\documentclass[aps,nofootinbib,preprintnumbers,twocolumn,superscriptaddress]{revtex4-1}

\usepackage{amssymb,amsmath,bm,natbib}
\usepackage{color}
\usepackage{slashed}
\usepackage{graphics}
\usepackage{graphicx}
\usepackage[dvipsnames]{xcolor}
\usepackage[utf8]{inputenc}
\usepackage[colorlinks = true,
            linkcolor = Maroon,
            urlcolor  = Maroon,
            citecolor = Maroon]{hyperref}
\usepackage{url}
\usepackage{dsfont}
\usepackage{float} 
\usepackage{cancel}
\usepackage{mhchem}
\usepackage{microtype}
\usepackage{soul}
\usepackage{xspace}
\usepackage{siunitx}
\usepackage{isotope}
\usepackage{lineno}
\usepackage{soul}

\newcommand{\TT}{\ensuremath{{^3\text{H}}}\xspace}
\newcommand{\He}{\ensuremath{{^3\text{He}^+}}\xspace}
\newcommand{\angstrom}{\textup{\AA}\xspace}

\begin{document}

\title{Heisenberg's uncertainty principle in the PTOLEMY project: a theory update}

\author{A.~Apponi}
\affiliation{INFN Sezione di Roma 3, Roma, Italy}
\affiliation{Universit\`a  di Roma Tre, Roma, Italy}

\author{M.~G.~Betti}
\affiliation{INFN Sezione di Roma 1, Roma, Italy}
\affiliation{Sapienza Universit\`a  di Roma, Roma, Italy}

\author{M.~Borghesi}
\affiliation{INFN Sezione di Milano-Bicocca, Milan, Italy}
\affiliation{Universit\`a di Milano-Bicocca, Milan, Italy}

\author{A.~Boyarsky}
\affiliation{Instituut-Lorentz, Universiteit Leiden, Leiden, The Netherlands}

\author{N.~Canci}
\affiliation{INFN Laboratori Nazionali del Gran Sasso, L'Aquila, Italy}

\author{G.~Cavoto}
\affiliation{INFN Sezione di Roma 1, Roma, Italy}
\affiliation{Sapienza Universit\`a  di Roma, Roma, Italy}

\author{C.~Chang}
\affiliation{Argonne National Laboratory, Chicago, IL, USA}
\affiliation{Kavli Institute for Cosmological Physics, University of Chicago, Chicago, IL, USA}

\author{V.~Cheianov}
\affiliation{Instituut-Lorentz, Universiteit Leiden, Leiden, The Netherlands}

\author{Y.~Cheipesh}
\affiliation{Instituut-Lorentz, Universiteit Leiden, Leiden, The Netherlands}

\author{W.~Chung}
\affiliation{Princeton University, Princeton, NJ, USA}

\author{A.~G.~Cocco}
\affiliation{INFN Sezione di Napoli, Napoli, Italy}

\author{A.~P.~Colijn}
\affiliation{Nationaal instituut voor subatomaire fysica (NIKHEF), Amsterdam, The Netherlands}
\affiliation{University of Amsterdam, Amsterdam, The Netherlands}

\author{N.~D'Ambrosio}
\affiliation{INFN Laboratori Nazionali del Gran Sasso, L'Aquila, Italy}

\author{N.~de~Groot}
\affiliation{Radboud University, Nijmegen, The Netherlands}

\author{A.~Esposito}
\affiliation{School of Natural Sciences, Institute for Advanced Study, Princeton, NJ, USA}

\author{M.~Faverzani}
\affiliation{INFN Sezione di Milano-Bicocca, Milan, Italy}
\affiliation{Universit\`a di Milano-Bicocca, Milan, Italy}

\author{A.~Ferella}
\affiliation{INFN Laboratori Nazionali del Gran Sasso, L'Aquila, Italy}
\affiliation{Universit\`a di L'Aquila, L'Aquila, Italy}

\author{E.~Ferri}
\affiliation{INFN Sezione di Milano-Bicocca, Milan, Italy}

\author{L.~Ficcadenti}
\affiliation{INFN Sezione di Roma 1, Roma, Italy}
\affiliation{Sapienza Universit\`a  di Roma, Roma, Italy}

\author{T.~Frederico}
\affiliation{Instituto Tecnol\'ogico de Aeron\'autica, S\~ao Jos\'e dos Campos, Brazil}

\author{S.~Gariazzo}
\affiliation{INFN Sezione di Torino, Torino, Italy}

\author{F.~Gatti}
\affiliation{Universit\`a di Genova e INFN Sezione di Genova, Genova, Italy}

\author{C.~Gentile}
\affiliation{Princeton Plasma Physics Laboratory, Princeton, NJ, USA}

\author{A.~Giachero}
\affiliation{INFN Sezione di Milano-Bicocca, Milan, Italy}
\affiliation{Universit\`a di Milano-Bicocca, Milan, Italy}

\author{Y.~Hochberg}
\affiliation{Racah Institute of Physics, Hebrew University of Jerusalem, Jerusalem, Israel}

\author{Y.~Kahn}
\affiliation{Kavli Institute for Cosmological Physics, University of Chicago, Chicago, IL, USA}
\affiliation{University of Illinois Urbana-Champaign, Urbana, IL, USA}

\author{M.~Lisanti}
\affiliation{Princeton University, Princeton, NJ, USA}

\author{G.~Mangano}
\affiliation{INFN Sezione di Napoli, Napoli, Italy}
\affiliation{Universit\`a degli Studi di Napoli Federico II, Napoli, Italy}

\author{L.~E.~Marcucci}
\affiliation{INFN Sezione di Pisa, Pisa, Italy}
\affiliation{Universit\`a degli Studi di Pisa, Pisa, Italy}

\author{C.~Mariani}
\affiliation{INFN Sezione di Roma 1, Roma, Italy}
\affiliation{Sapienza Universit\`a  di Roma, Roma, Italy}

\author{M.~Marques}
\affiliation{Instituto Tecnol\'ogico de Aeron\'autica, S\~ao Jos\'e dos Campos, Brazil}

\author{G.~Menichetti}
\affiliation{Universit\`a degli Studi di Pisa, Pisa, Italy}
\affiliation{Center for Nanotechnology Innovation @NEST, Istituto Italiano di Technologia, Pisa, Italy}

\author{M.~Messina}
\affiliation{INFN Laboratori Nazionali del Gran Sasso, L'Aquila, Italy}

\author{O.~Mikulenko}
\affiliation{Instituut-Lorentz, Universiteit Leiden, Leiden, The Netherlands}

\author{E.~Monticone}
\affiliation{INFN Sezione di Torino, Torino, Italy}
\affiliation{Istituto Nazionale di Ricerca Metrologica (INRiM), Torino, Italy}

\author{A.~Nucciotti}
\affiliation{INFN Sezione di Milano-Bicocca, Milan, Italy}
\affiliation{Universit\`a di Milano-Bicocca, Milan, Italy}

\author{D.~Orlandi}
\affiliation{INFN Laboratori Nazionali del Gran Sasso, L'Aquila, Italy}

\author{F.~Pandolfi}
\affiliation{INFN Sezione di Roma 1, Roma, Italy}

\author{S.~Parlati}
\affiliation{INFN Laboratori Nazionali del Gran Sasso, L'Aquila, Italy}

\author{C.~Pepe}
\affiliation{INFN Sezione di Torino, Torino, Italy}
\affiliation{Istituto Nazionale di Ricerca Metrologica (INRiM), Torino, Italy}
\affiliation{Dipartimento di Elettronica e Telecomunicazioni, Politecnico di Torino, Torino, Italy}

\author{C.~P\'erez~de~los~Heros}
\affiliation{Uppsala University, Uppsala, Sweden}

\author{O.~Pisanti}
\affiliation{INFN Sezione di Napoli, Napoli, Italy}
\affiliation{Universit\`a degli Studi di Napoli Federico II, Napoli, Italy}

\author{M.~Polini}
\affiliation{Universit\`a degli Studi di Pisa, Pisa, Italy}
\affiliation{Istituto Italiano di Tecnologia, Graphene Labs, Genova, Italy}
\affiliation{University of Manchester, Manchester, United Kingdom}

\author{A.~D.~Polosa}
\affiliation{INFN Sezione di Roma 1, Roma, Italy}
\affiliation{Sapienza Universit\`a  di Roma, Roma, Italy}

\author{A.~Puiu}
\affiliation{INFN Laboratori Nazionali del Gran Sasso, L'Aquila, Italy}
\affiliation{Gran Sasso Science Institute (GSSI), L'Aquila, Italy}

\author{I.~Rago}
\affiliation{INFN Sezione di Roma 1, Roma, Italy}
\affiliation{Sapienza Universit\`a  di Roma, Roma, Italy}

\author{Y.~Raitses}
\affiliation{Princeton Plasma Physics Laboratory, Princeton, NJ, USA}

\author{M.~Rajteri}
\affiliation{INFN Sezione di Torino, Torino, Italy}
\affiliation{Istituto Nazionale di Ricerca Metrologica (INRiM), Torino, Italy}

\author{N.~Rossi}
\affiliation{INFN Laboratori Nazionali del Gran Sasso, L'Aquila, Italy}

\author{K.~Rozwadowska}
\affiliation{INFN Laboratori Nazionali del Gran Sasso, L'Aquila, Italy}
\affiliation{Gran Sasso Science Institute (GSSI), L'Aquila, Italy}

\author{I.~Rucandio}
\affiliation{Centro de Investigaciones Energ\'eticas, Medioambientales y Tecnol\'ogicas (CIEMAT), Madrid, Spain}

\author{A.~Ruocco}
\affiliation{INFN Sezione di Roma 3, Roma, Italy}
\affiliation{Universit\`a  di Roma Tre, Roma, Italy}

\author{C.~F.~Strid}
\affiliation{Johannes Gutenberg-Universität Mainz, Germany}

\author{A.~Tan}
\affiliation{Princeton University, Princeton, NJ, USA}

\author{L.~K.~Teles}
\affiliation{Instituto Tecnol\'ogico de Aeron\'autica, S\~ao Jos\'e dos Campos, Brazil}

\author{V.~Tozzini}
\affiliation{Istituto Nanoscienze--CNR, NEST--Scuola Normale Superiore, Pisa, Italy}

\author{C.~G.~Tully}
\affiliation{Princeton University, Princeton, NJ, USA}

\author{M.~Viviani}
\affiliation{INFN Sezione di Pisa, Pisa, Italy}

\author{U.~Zeitler}
\affiliation{Radboud University, Nijmegen, The Netherlands}

\author{F.~Zhao}
\affiliation{Princeton University, Princeton, NJ, USA}

\collaboration{PTOLEMY Collaboration}


\begin{abstract}
We discuss the consequences of the quantum uncertainty on the spectrum of the electron emitted by the $\beta$-processes of a tritium atom bound to a graphene sheet. We analyze quantitatively the issue recently raised in~\cite{Cheipesh:2021fmg}, and discuss the relevant time scales and the degrees of freedom that can contribute to the intrinsic spread in the electron energy. We perform careful calculations of the potential between tritium and graphene with different coverages and geometries. With this at hand, we propose possible avenues to mitigate the effect of the quantum uncertainty. 
\end{abstract}

\maketitle


\section{Introduction}

Neutrinos are one of the most elusive particles known to us, and many questions regarding their nature remain unanswered. On the one hand, it is now well assessed that at least two of the three standard neutrinos are massive~\cite{Super-Kamiokande:1998kpq,SNO:2001kpb,SNO:2002tuh,deSalas:2020pgw,Esteban:2020cvm,Capozzi:2021fjo}, and the values of their squared mass differences are known with good precision (see, e.g.,~\cite{deSalas:2017kay,Capozzi:2018ubv,Capozzi:2017ipn}). On the other hand, we still do not know their absolute mass scale (i.e. the mass of the lightest neutrino) and mass ordering (i.e. whether the lightest neutrino is mostly within the first or third leptonic family). Moreover, the existence of a cosmic neutrino background is a robust prediction of the current cosmological paradigm, and it is expected to carry a plethora of information about the early stages of the Universe~\cite{Lesgourgues:2013sjj}. Despite many indirect evidences, it has however yet to be observed directly.

One of the best ways to determine the absolute neutrino mass scale is by studying the spectrum of the electron emitted by $\beta$-decay close to its maximum kinetic energy, the so-called endpoint. The best up to date bound on the effective neutrino mass, $m_\nu^2 = \sum_i {|U_{ei}|}^2 m_i^2$~\cite{Formaggio:2021nfz}, is $m_\nu < 0.8$~eV at $90\%$ C.L., as obtained by the KATRIN experiment using gaseous molecular tritium~\cite{KATRIN:2019yun,KATRIN:2021uub}. As far as the cosmic neutrino background is concerned, instead, one could in principle detect it via the process of neutrino capture~\cite{Cocco:2007za,Li:2010sn,Faessler:2013jla,Long:2014zva,Betti:2018bjv,Abdelnabi_2020,Apponi:2021hdu,Apponi_2021,nano11010130}. In this case, if the process happens in vacuum, the energy of the emitted electron is expected to be larger than the endpoint by twice $m_\nu$. However, a finite experimental resolution broadens the observed spectrum, turning the $\beta$-decay contribution into the main source of background, which could hide the absorption peak.

The proposed PTOLEMY experiment~\cite{PTOLEMY:2019hkd} is expected to address both points. In particular, the goal is to study the spectrum of electrons produced by the decay and absorption processes of atomic tritium:\footnote{For the possibility of using a heavier emitter with an endpoint similar to tritium, see~\cite{deGroot:2022tbi}.}
\begin{subequations} \label{eq:processes}
    \begin{align}
        \TT &\to \He + e^- + \bar \nu_e\,, \label{eq:decay} \\
        \nu_e + \TT &\to \He + e^-\,. \label{eq:capture}
    \end{align}
\end{subequations}
The proposed target substrate is graphene~\cite{PTOLEMY:2019hkd}, which can efficiently store atomic tritium by locally binding it to carbon atoms, hence allowing a large target mass in a small scale experiment, and providing a good voltage reference for an electromagnetic spectrometer. Thanks to the corresponding large event rate, this should substantially improve on the existing bounds for the absolute mass scale, by accurately measuring the normalization of the spectrum near the end-point, which is sensitive to the lightest neutrino mass. Indeed, it is expected for PTOLEMY to have the sensitivity to measure an effective mass as small as $m_\nu=50$~meV already at the early stages of the experiment, with a target mass of 10~mg~\cite{PTOLEMY:2019hkd}. This is almost completely independent on the experimental energy resolution.

To detect cosmic neutrinos from the emitted electron spectrum, instead, one needs to resolve the peaks coming from the capture process~\eqref{eq:capture} from the contribution coming from the standard $\beta$-decay~\eqref{eq:decay}. To do that, both a large event rate and a precise determination of the electron energy are required. For the former, a much larger target mass is required and PTOLEMY would ideally have 100~g of tritium, while for the latter, it is expected to achieve an energy resolution as small as $100$~meV~\cite{PTOLEMY:2019hkd}.

Nonetheless, as first pointed in~\cite{Cheipesh:2021fmg} and recently reanalyzed in~\cite{Nussinov:2021zrj}, since a condensed matter substrate is implicitly present in the processes of Eqs.~\eqref{eq:processes}, Heisenberg's principle implies an intrinsic uncertainty in the electron energy. Indeed, since the initial tritium is spatially confined, it has an associated spread in its momentum, which in turns propagates to a spread in the energy of the emitted electron. Using the graphene--tritium binding potentials available in literature~\cite{moaied2015theoretical,gonzalez2019hydrogen}, this spread is expected to be about an order of magnitude larger than the desired experimental resolution. Consequently, to still be able to determine whether or not the absorption process has been detected, one would need sufficiently accurate theoretical predictions for the electron spectrum, along the lines of what was done for the KATRIN experiment~\cite{PhysRevLett.77.4724,saenz1997effect,Saenz:2000dul,Doss:2007xia,Bodine:2015sma}. This, in turns, requires a detailed knowledge of the initial and final state of the reaction. Due to the intricacies of the condensed matter state, this is a task substantially harder than originally expected.

In this work we provide a theoretical update on the issue reported above. This is done by both spelling out the problem in a firm quantitative way, as well as by proposing possible avenues to mitigate it. First, we quantitatively review the issue, with emphasis on the role played by different final states of the reaction. We also propose an experimental test to verify the understanding behind the predicted spread in the electron energy. We then discuss the relevant degrees of freedom that could influence the spectrum of the emitted electron, both before and after the reaction. Possible avenues to circumvent the problem are also discussed. In particular, a solution to this issue should likely come from a judicious tuning of the initial state for the reactions in Eq.~\eqref{eq:processes}, rather that from the inclusion of further degrees of freedom in the final state. We present possible ways to engineer the initial tritium wave function using geometries alternative to flat graphene.

\vspace{1em}

\noindent \emph{Conventions:} Throughout this manuscript we work in natural units, $\hbar = c = k_B =1$.


\section{Quantum spread in a nutshell} \label{sec:nutshell}

In a system hosting a plethora of different degrees of freedom, like the one at hand, there are different effects that can modify the spectrum of the outgoing electron. These can be schematically classified as follows:
\begin{enumerate}
    \item \emph{Initial state effects.} These concern the state of the system before the $\beta$-processes take place. One such effect is related to the fact that, in general, the decaying atom will not be in an eigenstate of momentum. The associated initial momentum spread will affect the energy of the outgoing electron---see~\cite{Cheipesh:2021fmg}. Moreover, quasi-particle degrees of freedom (e.g., phonons), present before the decay, will affect the experimental resolution through their thermal and zero-point motion, which contribute to the broadening of the initial tritium momentum.
    \item \emph{Final state effects.} These, instead, concern the state of the system as it is left after the $\beta$-processes. Specifically, radioactive decay may create one or several quasi-particle excitations such as a phonons, electron-hole pairs or plasmons. Each of them subtracts a fraction of energy from the electron and, being indistinguishable instances, causes further broadening.
\end{enumerate}
In this section we focus on the effects due to the spatial localization of the initial tritium wave function. We briefly review the key points of the argument formulated in~\cite{Cheipesh:2021fmg}, as well as compute the expected electron rate, under certain simplifying assumptions. For the sake of the current argument, it suffices to consider the case of a single neutrino. 

Let us work at finite volume, $V$, and consider an initial \TT atom with wave function $\psi_i(\bm{x}_\text{T})$, and a final \He with wave function $\psi_f(\bm{x}_\text{He})$. As we will explain in detail, the effect discussed here is a sole consequence of the spatial localization of the initial wave function. For this reason, it is enough to assume all the initial tritium atoms to be in the ground states. Instead, both the electron and neutrino wave functions are plane waves, $\psi_\beta(\bm{x}_\beta) = e^{i \bm{k}_\beta\cdot\bm{x}_\beta}/\sqrt{V}$ and $\psi_\nu(\bm{x}_\nu) = e^{i \bm{k}_\nu\cdot\bm{x}_\nu}/\sqrt{V}$. On large enough distances, the weak interaction Hamiltonian is roughly constant, and the location of the final decay products is the same as the initial tritium.
%
%
This means that the transition matrix element from the initial state, $i$, to a specific final state, $f$, is\footnote{The coupling $g$ can be related to the microscopic theory of weak interactions by~\cite{Simkovic:2007yi}
\begin{align*}
    {|g|}^2 = G_{F}^2 \big|V_{ud}\big|^2  \left( f_{V}^2 \big| \mathcal{M}_{F} \big|^2 + f_A^2 \big| \mathcal{M}_{GT} \big|^2 \right)\,,
\end{align*}
with $G_{F}$ the Fermi constant, $V_{ud}$ the entry of the CKM matrix, $f_V$ and $f_A$ the vector and axial couplings of the nucleon, and $\mathcal{M}_F$ and $\mathcal{M}_{GT}$ the so-called Fermi and Gamow--Teller matrix elements. We are neglecting the relativistic Coulomb form factor for simplicity, since it is approximately one. Here we use $f_V=1$, $f_A=1.25$, $|\mathcal{M}_F|=1$~\cite{Simkovic:2007yi} and $|\mathcal{M}_{GT}|=1.65$~\cite{Baroni:2016xll}.}
\begin{align} \label{eq:M}
    \mathcal{M}_{fi} = \frac{g}{V} \int d^3x \, \psi_i(\bm{x}) \, \psi^*_f(\bm{x}) \,e^{-i(\bm{k}_\beta+\bm{k}_\nu)\cdot \bm{x}}\,.
\end{align}
Using Fermi's golden rule, the corresponding transition probability rate is
\begin{align}
    d\Gamma_{fi} = (2\pi) \big|\mathcal{M}_{fi}\big|^2 \delta(E_i - E_f - E_\beta - E_\nu) d\rho_{f}\,,
\end{align}
where $E_i$ is the energy of the initial bound tritium (rest mass + negative binding energy), $E_f$ the energy of the final helium (depending on the particular final state under consideration), and $E_\beta$ and $E_\nu$ the relativistic electron and neutrino energies. Moreover, $d\rho_f$ is the phase space of the final decay products, which also depends on the particular final state. 

For the typical graphene--tritium potential, the initial wave function is roughly Gaussian,
\begin{align}
    \psi_i(\bm{x}) = \frac{1}{\pi^{3/4}\lambda^{3/2}} e^{-\frac{x^2}{2\lambda^2}} \,,
\end{align}
localized in space within a distance $\lambda$, which is of order of a few fractions of an \angstrom, but whose precise value depends on the details of the substrate.\footnote{For the sake of the current argument, we consider an isotropic case, where all directions are equivalent, as discussed in~\cite{Cheipesh:2021fmg} In a more realistic scenario $\lambda$ would be a matrix---see also Section~\ref{sec:potentials}. For an isotropic harmonic potential, the spread of the wave function is related to the tritium mass, $m_\TT$, and to the spring constant, $\kappa$, by $\lambda \equiv \left(m_\TT \kappa\right)^{-1/4}$.}
Since the initial state is not a momentum eigenstate, exact momentum conservation is spoiled: it will only be satisfied up to a spread of order $\sim 1/\lambda$. 

Two qualitatively different scenarios can now arise.\footnote{We focus here on the two extreme scenarios. An infinite number of possible final states will interpolate between these two instances.} On the one hand, the \He might remain bound to the graphene, ending up in some discrete level of its potential. In particular, on the short time scales over which the $\beta$-processes happen, it will be subject to the same binding potential as the initial tritium, as explained in Section~\ref{sec:scales}. Among all these final states, when the \He remains in the ground state is when the outgoing electron can have kinetic energy as high as possible. For events where also no lattice vibrational modes are excited, this corresponds to $K_\beta^\text{max}=Q-m_\nu$, where $Q\equiv m_
\TT - m_\He - m_e\simeq 18.6$~keV~\cite{PhysRevLett.70.2888} is the $Q$-value in vacuum.\footnote{Here $m_\TT$ and $m_\He$ are the atomic masses, see also~\cite{PTOLEMY:2019hkd}.} However, in this scenario the final helium wave function, $\psi_f$, is itself localized in space, and the corresponding matrix element, near the endpoint reads,
\begin{align}
    \begin{split}
        \mathcal{M}_{fi} \simeq{}& \frac{g}{V} \int d^3x \Bigg( \frac{e^{-\frac{x^2}{2\lambda^2}}}{\pi^{3/4}\lambda^{3/2}} \Bigg)^2
        e^{-i \bm{k}_\beta\cdot\bm{x}} \\
        ={}& \frac{g}{V} e^{-\lambda^2 k_\beta^2/4} \,.
    \end{split}
\end{align}
%
Due to the large separation of scales between the electron momentum and the typical atomic size, this is exponentially suppressed. In particular, for flat graphene at maximum coverage, near the endpoint one has $\lambda k_\beta\simeq 6$. This makes the events close to the endpoint (corresponding to electrons with highest possible momentum) extremely unlikely. This instance is completely analogous to what happens in the M\"ossbauer effect (see, e.g.,~\cite{lipkin1960some}).

In the second scenario the \He is freed (or almost freed) from the graphene sheet, i.e. it is excited close to or above the absolute zero of the potential. In this case, the maximum electron kinetic energy is $Q - \varepsilon_0 - m_\nu$, with $\varepsilon_0$ the ground state binding energy of the initial tritium. The matrix element near the endpoint is now given by,
\begin{align}
    \begin{split}
        \mathcal{M}_{fi} \simeq{}& \frac{g}{V^{3/2}} \int d^3x \, \frac{e^{-\frac{x^2}{2\lambda^2}}}{\pi^{3/4}\lambda^{3/2}} \, e^{-i (\bm{k}_\text{He} + \bm{k}_\beta) \cdot \bm{x}} \\
        ={}& \frac{g}{V^{3/2}} 2^{3/2} \pi^{3/4} \lambda^{3/2}e^{-\lambda^2|\bm{k}_\text{He} + \bm{k}_\beta |^2/2} \,.
    \end{split}
\end{align}
%
Close to the maximum energy allowed by the process, this is still exponentially suppressed but, moving to smaller energies, the momentum of the outgoing helium can compensate that of the electron, $\bm{k}_\text{He}\simeq-\bm{k}_\beta$, making the probability for this final state sizable. Very similar arguments hold for the rate of absorption of a cosmic neutrino. 

The result is an overall distortion of the electron spectrum, which not only changes its emitted energy, but it also makes the absorption peaks either disappear under the $\beta$-decay part of the spectrum or extremely rare. In Figure~\ref{fig:rates} we quantify the above observations. Further details for the calculation are reported in Appendix~\ref{app:rates}. 

Note that, when no vibrational mode is excited, the recoil energy of graphene is completely negligible due to its large mass. The electrons emitted in this instance (red lines in Figure~\ref{fig:rates}) are therefore more energetic than the endpoint in vacuum. The maximum allowed energy is larger precisely by an amount equal to the recoil energy that the \He would have in vacuum, $K_\text{rec}\simeq 3.44$~eV.\footnote{A similar effect should also be present in the KATRIN setup~\cite{Bodine:2015sma}. In that case, the best one can do is to transmit the momentum to the molecule as a whole. With respect to vacuum, this should increase the maximum electron energy by roughly $\frac{m_\TT}{m_\TT+m_\He}K_\text{rec}\simeq1.72$~eV.} This feature is unique to processes involving a condensed matter substrate in the initial state, which can absorb part of the recoil momentum. The observation of electrons with energy higher than the endpoint in vacuum would then constitute an experimental observable to probe initial state effects.

It is clear from Figure~\ref{fig:rates} that the rate for the ground state-to-ground state transition has a shape that would be ideal for cosmic neutrino background detection, with the absorption peak well separated from the decay continuum. It is, however, exponentially unlikely. One would then like to maximize its probability, which could be achieved by making the initial state tritium as close as possible to a momentum eigenstate, i.e. as delocalized as possible. This way, the dominant process would feature a free particle both in the initial and final state, which maximizes the probability while decreasing the quantum spread. One potential solution in this direction is discussed in Section~\ref{sec:nanotubes}.

\begin{figure}
    \centering
    \includegraphics[width=\columnwidth]{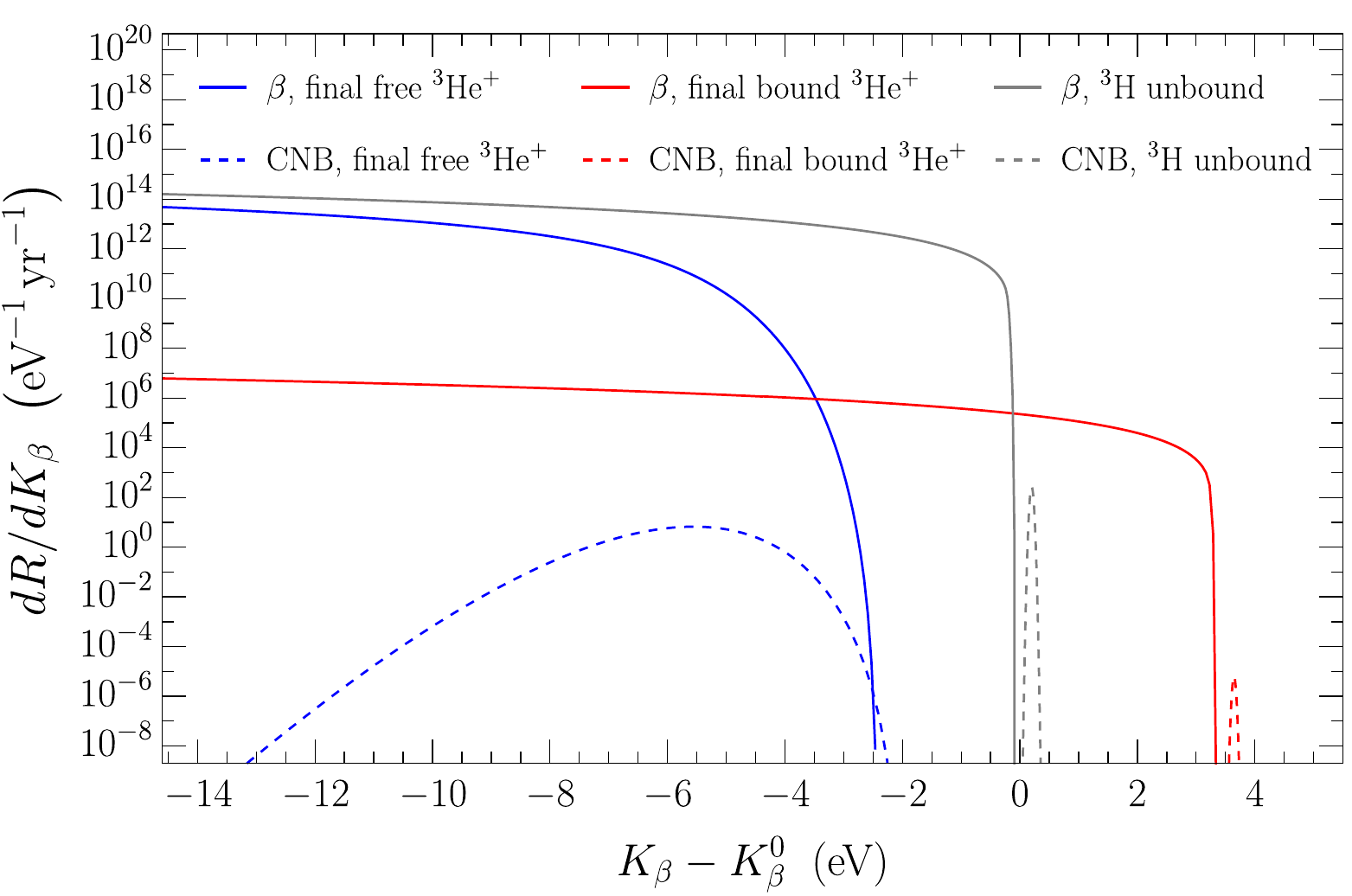}
    \caption{Event rates for different configurations as a function of the outgoing electron kinetic energy, measured with respect to the endpoint in vacuum at zero neutrino mass, $K_\beta^0 \equiv m_\TT - m_\He - m_e - K_\text{rec}$, with $K_\text{rec}\simeq3.44$~eV. We assume a target mass of 100~g, corresponding to $N_\text{T}\simeq 2\times 10^{25}$ atoms of tritium. The presence of the graphene in the initial state shifts the event rate to different energies and makes the absorption peaks (CNB) much rarer, and hidden under the $\beta$-decay part of the spectrum. For illustrative purposes we have set $m_\nu = 0.2$~eV,  taken the initial wave function to be the ground state of the full coverage graphane potential presented later in Section~\ref{sec:potentials}, and convoluted with an experimental resolution described by a Gaussian with full width at half maximum $\Delta = 0.05$~eV. Moreover, we only considered the two extreme scenarios: the free helium (solid and dashed blue lines) and the helium bound to the ground state with no emission of vibrational modes (solid and dashed red lines). Intermediate instances will populate the regions between the blue and red lines, resulting in a smooth total rate. For comparison, we report the rate expected if the process were to happen in vacuum, i.e. with an initial free atomic \TT (gray lines).}
    \label{fig:rates}
\end{figure}


\section{Time scales, length scales and relevant degrees of freedom} \label{sec:scales}

Let us now qualitatively discuss the role of additional degrees of freedom and the interplay between the different scales present in the problem at hand. This discussion will partially overlap with what was presented in~\cite{Nussinov:2021zrj}.


As mentioned in Section~\ref{sec:nutshell}, initial state effects are due both to the particular state in which the tritium atom is, as well as to the presence of additional degrees of freedom, most notably, phonons. Consider the situation of a tritium atom chemically attached to a carbon atom in a free-standing graphene, as also discussed in~\cite{Nussinov:2021zrj}. Since the tritium is four times lighter than the carbon, we can neglect its effect on the vibrational spectrum of the lattice. The velocity of the tritium atom can be written as $\bm{v}_\text{T} = \bm{v}_\text{C} + \bm{v}_\text{TC}$,
where $\bm{v}_\text{C}$ is the velocity of the carbon, and $\bm{v}_\text{TC}$ that of the tritium relative to the 
carbon to which it is attached. The uncertainty on the tritium velocity then reads
\begin{align}
    {(\Delta \bm{v}_\text{T})}^2 = {(\Delta \bm{v}_\text{C})}^2+{(\Delta \bm{v}_\text{TC})}^2\,.
\end{align}
Here $\Delta \bm{v}_\text{TC}$ is the uncertainty in the tritium velocity due to the localization of its 
wave function near a carbon atom with which it forms a chemical 
bond (see the previous section). We have neglected correlations 
between the zero-point vibrations of the tritium--carbon bond and
zero-point motion of the carbon to which the tritium is attached (such an approximation can be justified along the lines of the Born--Oppenheimer approximation~\cite{sutcliffe2012quantum}). The effect of the initial state phonons in graphene is encoded in the thermal energy contribution (see, e.g.,~\cite{GRIMVALL1999112}), given by
\begin{align}
    {(\Delta \bm{v}_\text{C})}^2= \sum_{\bm{Q}} \sum_{s} \frac{\omega_{\bm{Q},s}}{2 m_\text{C}  N}\coth\left(\frac{\omega_{\bm{Q},s}}{2 T}\right)\,,
\end{align}
with $m_\text{C}$ the mass of a carbon atom, $N$ the number of carbon atoms in a sample, and  
$\bm{Q}$ the Bloch momentum of the graphene phonon. Moreover, $s$ enumerates the phonon polarization branch, and $\omega_{\bm{Q}, s}$ its dispersion relation. For temperatures larger than the Debye temperature (the bandwidth of the phonons), the phonon contribution reduces to the Dulong--Petit law, $(\Delta \bm{v}_\text{C})^2 =T/ m_\text{C}.$ The contribution decreases steadily with decreasing temperature reaching, in the zero temperature limit, its intrinsic quantum value, $(\Delta \bm{v}_\text{C})^2 = \bar{\omega}/2m_\text{C},$ where $\bar{\omega}$ is the Brillouin-zone average phonon frequency, which is on the same order as the phonon bandwidth. The initial state vibrational modes will then contribute to the broadening of the tritium velocity, $\Delta\bm{v}_\text{T}$. This phenomenon is, however, suppressed with respect to $\Delta \bm{v}_\text{TC}$ by factor $m_\TT/m_\text{C}$, causing an order $\sim 10$~\% effect---see also~\cite{Nussinov:2021zrj}.

The second class of effects, the final state ones, leads to the broadening of the spectrum due to the following reason: each quasi-particle will generally carry away some amount of energy bounded by its bandwidth. 
Even focusing on the softest quasi-particle, the flexural phonons~\cite{jiang2015review}, the branch for this channel has a span of $\sim 0.1$~eV. This ensures that the creation of one such quasi-particle leads to the loss of required energy resolution. 

What degrees of freedom can contribute to final state broadening? The key quantity to consider is the time over which the $\beta$-decay ``happens'', since this is when the emitted electron decouples from the rest of the system. This time scale can be determined from the formation time of the process, i.e. from the time it takes to separate the electron wave function from the helium one. Close to the endpoint the electron carries a momentum $k_\beta\simeq 139$~keV. Its de Broglie wavelength is then roughly 0.01~\angstrom, making it point-like compared to the \He atom, whose radius is $r_\text{He}\sim1~\angstrom$. The formation time is then simply given by
\begin{align}
    t_\beta \sim \frac{m_e r_\text{He}}{k_\beta} \sim 10^{-18}~\text{s}\,.
\end{align}
Final state degrees of freedom which are excited over time scales significantly larger than this are essentially irrelevant for what concerns the spectrum of the observed electron, in agreement with the so-called sudden approximation (see also~\cite{Nussinov:2021zrj}). 
An example of degrees of freedom whose excitation happens over times scales shorter than $t_\beta$, and that could hence influence the final spectrum, are electrons located at small distances from the decay point, or the vibrational modes of the nearby carbon atom. Indeed, their excitation happens via the propagation of the Coulomb potential of the \He atom, which travels at the speed of light.

Along similar lines, since the $\beta$ electron velocity is much larger than both the Fermi velocity and then \He velocity, $v_\beta \gg v_F \gg v_\text{He}$, the helium atom and the electrons of the graphene can be considered as frozen on the positions they occupy when the reaction happens. In particular, this means that the \He atom will experience the same potential as the initial \TT. The change to the new potential, in fact, follows the rearrangement of the graphene electrons induced by the positive charge, which happens over times significantly larger then $t_\beta$ (see, again,~\cite{Nussinov:2021zrj}).\footnote{Since Coulomb interactions are long range, the rearranging graphene electrons and the \He ion, despite being slow, could actually play a role, by contributing to the corrections to the sudden approximation. Indeed, such a subleading effect has already proved to be of some relevance in~\cite{PhysRevLett.77.4724,saenz1997effect}. The inclusion of this effect is beyond the scope of the present work.}

\vspace{1em}

Let us conclude this section with a comment. One might wonder whether those degrees of freedom that can indeed be excited in the final state of the reaction (atomic electrons, lattice vibrations, etc.) can help alleviate the problem of the exponential suppression of the matrix element discussed in the previous section, by compensating for the electron momentum in other ways.
This is not the case, as we now argue. Consider a set of possible degrees of freedom, whose mass we represent with $m_i$. The total mass of the system after the reaction (excluding the $\beta$ electron) will be $M=m_\He + \sum_i m_i$. If we denote their positions as $\bm{x}_i$, the center of mass of the system and the distances from the helium atom are
\begin{align}
    \bm{R} = \frac{\bm{x}_\text{He}m_\He + \sum_i  \bm{x}_i m_i}{M}\,, \qquad \bm{r}_i = \bm{x}_\text{He} - \bm{x}_i\,.
\end{align}
After the decay, the plane wave of the $\beta$ electron, which is located at the same position as the helium atom, can hence be rewritten as
\begin{align}
    e^{i\bm{k}_\beta\cdot\bm{x}_\beta} = e^{i\bm{k}_\beta\cdot\bm{R}}\prod_i e^{i\frac{m_i}{M} \bm{k}_\beta\cdot\bm{r}_i}\,, \;\; \text{ since } \;\; \bm{x}_\beta = \bm{x}_\text{He}\,.
\end{align}
Assuming that the typical separation between the different components is of atomic size, $a \sim 1$~\angstrom, this tells us that light degrees of freedom for which $(m_i/M)k_\beta a\lesssim 1$ can be excited to discrete levels with sizable probability, since they will not suffer from the exponential suppression described in the previous section. Heavy degrees of freedom, instead, for which $(m_i/M)k_\beta a \gg 1$, must be liberated, or else the matrix element is strongly suppressed. Whenever this happens, a mechanism like the one explained in the previous section will cause an intrinsic spread in the electron energy. In Appendix~\ref{app:recoil}, we show this in the simple setting of atomic tritium.


\section{The substrate--tritium interaction potential} \label{sec:potentials}

As already explained, the most substantial contribution to the intrinsic uncertainty on the electron energy comes from the localization of the initial tritium wave function. To tackle the problem at hand, it is then crucial to have detailed knowledge of the initial state of the reactions. We therefore start by studying the interaction between hydrogen and graphene, given that the former has the same chemical properties of tritium. We then also consider different possible graphene derived materials. These could be alternatives to manipulate the tritium potential---i.e. the initial state of our reactions---and optimize it in order to mitigate the intrinsic quantum effects.


\subsection{Tritium on extended graphene}

Let us start by first assessing the form of the potential between hydrogen and extended graphene, similar to what was considered in~\cite{Cheipesh:2021fmg}. The binding of the hydrogen happens to a carbon site with a $C_3$ symmetry.
The  potential has two contributions, one perpendicular and one parallel to the sheet. The former is simply the binding potential, while the latter is the hopping potential, which controls the mobility of the atom along the surface of the graphene sheet. These potentials depend sensitively on the pristine status of graphene (i.e. its local structure, doping and hydrogen coverage), and several \emph{ab initio} studies are available in the literature. 

\begin{figure}[t]
    \centering
        \includegraphics[width=\columnwidth]{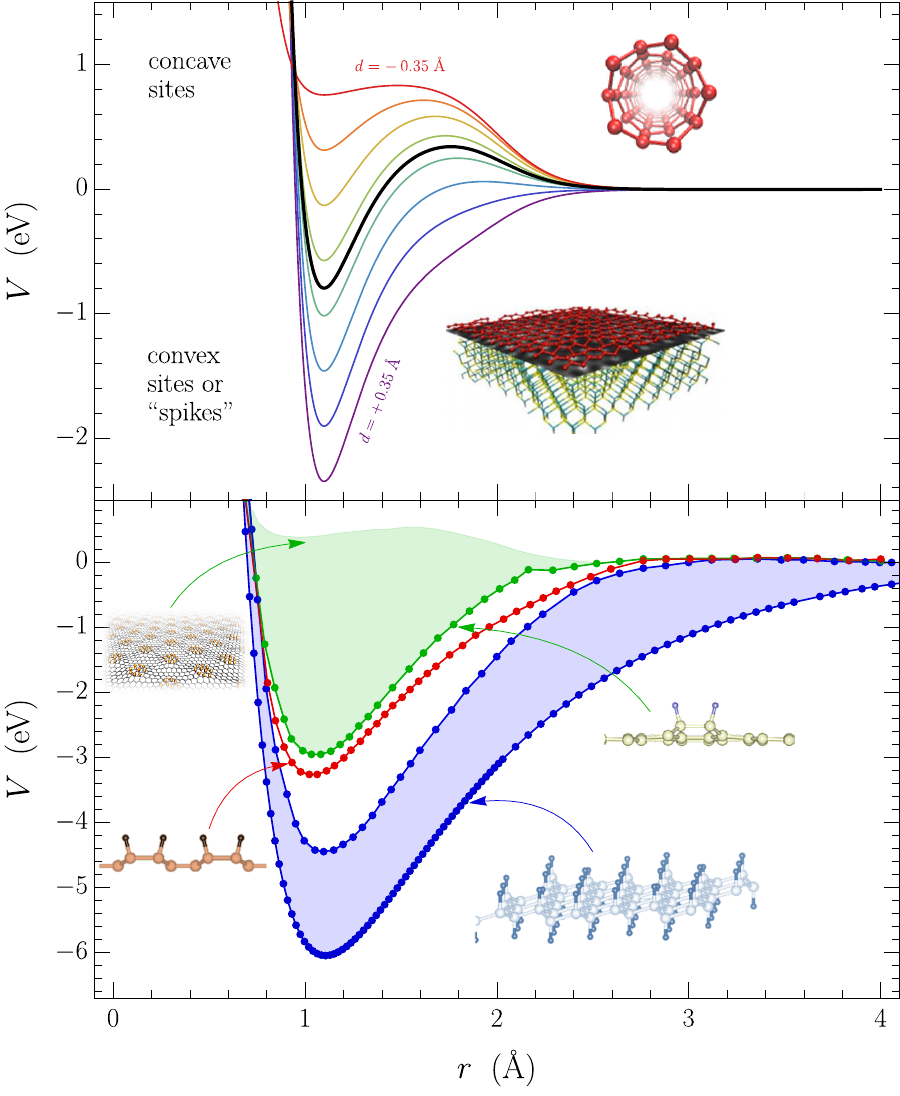} 
        \caption{Graphene--hydrogen binding potential as a function of the distance from the binding site. \textbf{Upper panel:} Binding potential for a single hydrogen atom for different local curvatures (puckering) of the binding site---see Figure~\ref{fig:pucker} for a definition. Flat graphene corresponds to $d=0$ (thick black line), while $d>0$ corresponds to convex sites (outward puckering, as in the spikes of crumpled supported sheets shown in the lower inset) and $d<0$ to concave ones (inward puckering, as within the nanotube shown in the upper inset).  The potentials are parametrized as an interpolation between a van der Waals potential at large distances and a binding potential at short distances. The depth of the latter depends on the curvature---see Appendix~\ref{app:valapp} for details---which has been varied between $d=-0.35$~\angstrom and $d=+0.35$~\angstrom. Tritium is expected to have the same chemical properties as hydrogen. \textbf{Lower panel:} Effect of the hydrogen coverage. Same side binding has a positive cooperative effect for dimers (red line) or cluster up to a small value of the coverage (green line). As the clusters size increases (i.e. when more atoms are added on the same side) the binding destabilizes due to mechanical distortion (green shade). Conversely, two sides coverage is more stable (blue lines) although still dependent on coverage (blue shade). Representative structures are reported. Energy profiles are obtained with a standard Density Functional Theory calculation, as reported in Appendix~\ref{app:dft}.}
    \label{fig:potentials}
\end{figure}

Consider first the case of a single isolated atom binding to the flat graphene. The binding energy with respect to the atomic state has been evaluated to be around $0.7-0.8$~eV, arising from a potential minimum located around $1.1$~\angstrom. There is then a barrier of $0.2-0.3$~eV, and a van der Waals well, with minimum at about $2.5-3$~\angstrom, and a very shallow depth estimated to be $5-7$~meV (see Figure~\ref{fig:potentials}, upper panel, thick black line)
~\cite{moaied2015theoretical,gonzalez2019hydrogen,Bonfanti_2018a,Bonfanti_2018b,Casolo_2009,PhysRevB.75.245413,ivanovskaya2010hydrogen,boukhvalov2010modeling,PhysRevB.77.035427,wang12}. It was also shown that the binding energy is strongly dependent on the local curvature of the sheet~\cite{PhysRevB.77.035427,valjcpc2011,Pizzochero_2015,Pizzochero_2016}. In particular, it increases up to an additional $1.5$~eV on very convex surfaces, as the exterior of small fullerenes~\cite{valpccp2013} or on ``spikes'' of crumpled carbon sheets forming on given substrates~\cite{golerJpcc2013}. Conversely, it decreases within concavities, as in the interior of carbon nanotubes. (See Figure~\ref{fig:potentials}, upper panel, colored lines~\cite{valjcpc2011}.)

Due to a cooperative effect, the hydrogen atoms have a tendency to dimerize (see  Figure~\ref{fig:potentials}, lower panel, green line) \cite{golerJpcc2013} on the graphene surface. The formation of large clusters of atoms bound on the same side, however, is limited by the consequent creation of curvature on the sheet, which destabilizes the structure hence decreasing the binding energy per atom (Figure~\ref{fig:potentials}, lower panel, green shade~\cite{rossijpcc2015}). In this respect, a more favorable high coverage setup is that of dimers separated by vacant sites (red line), bearing weak global curvature, or even graphane~\cite{science2009} (blue lines), where the sites of the triangular sublattice are occupied on different sides of the sheet. Even in this condition there is a dependence on the coverage of one side with respect to the other (blue shades), and the binding energies are in the range $4-6$~eV~\cite{prb75153401}.

The PTOLEMY proposal would like to achieve a high coverage of tritium, to maximize the event rate. In light of the considerations above, fully (half) occupied graphane are favorable conformations: these give a stoichiometry $\text{C}\!:\!\TT=1\!:\!1\,(2\!:\!1)$, meaning a $20\,(11)\%$ of gravimetric loading of tritium---$200\,(110)$~g of tritium per kg of material.

\begin{figure}
    \centering
    \includegraphics[width=\columnwidth]{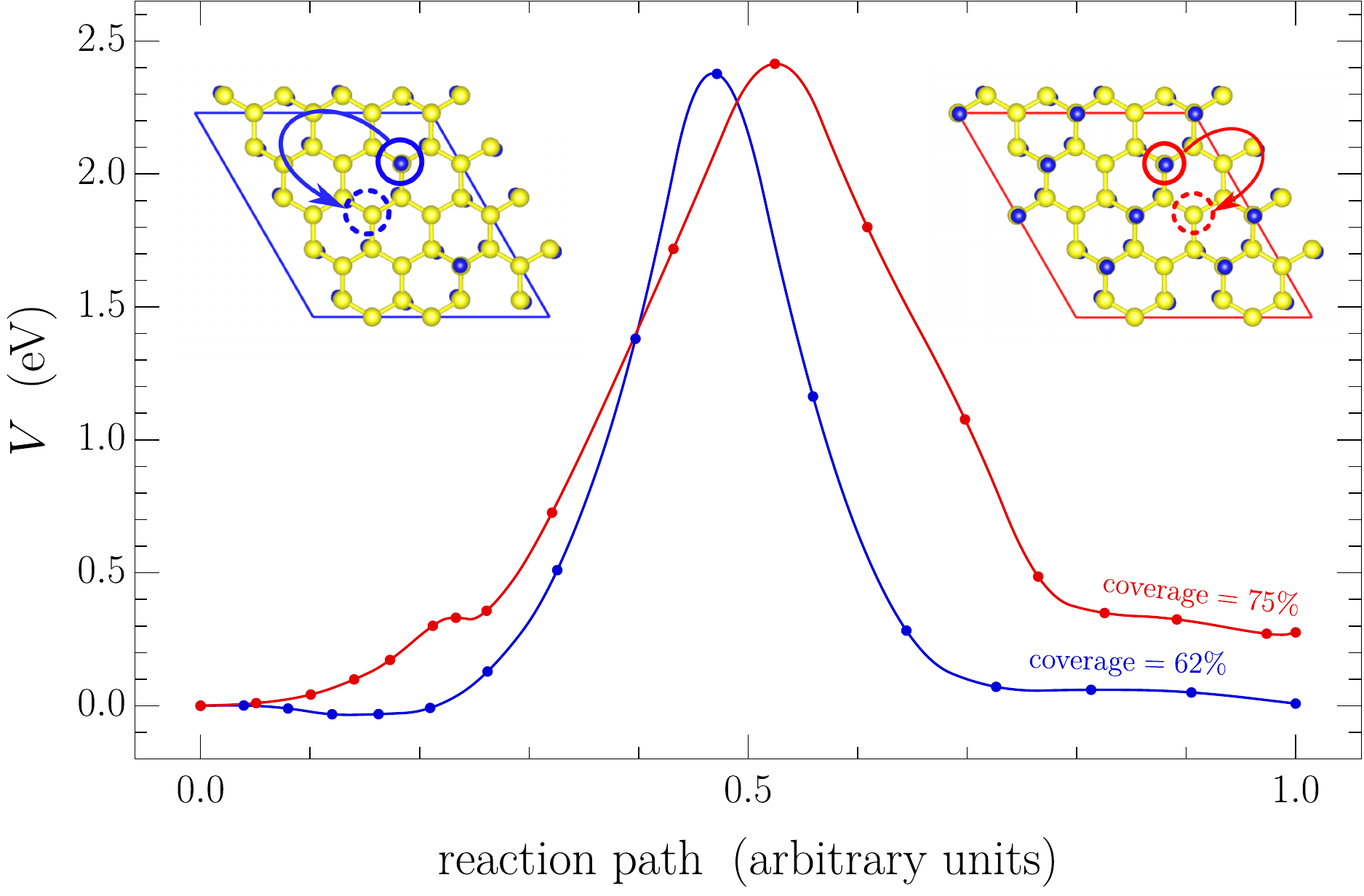}
    \caption{Hopping energy profiles on graphane at different coverage as a function of the reaction path (in arbitrary units), i.e. the path connecting the initial and final positions of the hydrogen atom. 
    In both examples one side is hydrogenated at alternating sites. In one instance the other side is only slightly hydrogenated, for a total coverage of 62\% (blue line, left inset), while in the other instance the hydrogenation is increased to a 75\% coverage (red line, right inset). In both cases, the vertical axis represents the energy required for a hydrogen atom to hop from a site to another. The two instances are represented in the insets, with blue dots being the hydrogen atoms and the arrows representing a typical hopping path.}
    \label{fig:hop}
\end{figure}

The hopping potential of hydrogen on graphene (i.e. its ability to move along the surface) is even less well characterized in the literature, and also likely to be dependent on the curvature and other local features of the sheet. As prototypical examples, we considered two cases with different coverage and local conformation (generated as described in Figure~\ref{fig:hop}). These represent both partially saturated graphane and the crumpled surface of supported graphene with partial covalent bonding to its substrate (see also Figure~\ref{fig:potentials}). We evaluated the hopping profiles within the framework of the Density Functional Theory (DFT, details reported in Appendix~\ref{app:dft}) finding a hopping barrier around $2$~eV. Its average value turns out lower than the average desorption energy (compare with Figure~\ref{fig:potentials}), and the profiles can be asymmetric, due to the irregular and disordered conformation of the hopping sites. Additionally, the minima between barriers appear shallower than in the desorption profile and modulated depending on the specific path. All this shows a dependence of the hopping barrier on the local geometry of the sheet, indicating the possible existence of specific paths along which the hopping, and hence the mobility of the tritium, is particularly favored. We now discuss one such possibility.

\begin{figure*}
    \centering
    \includegraphics[width=\textwidth]{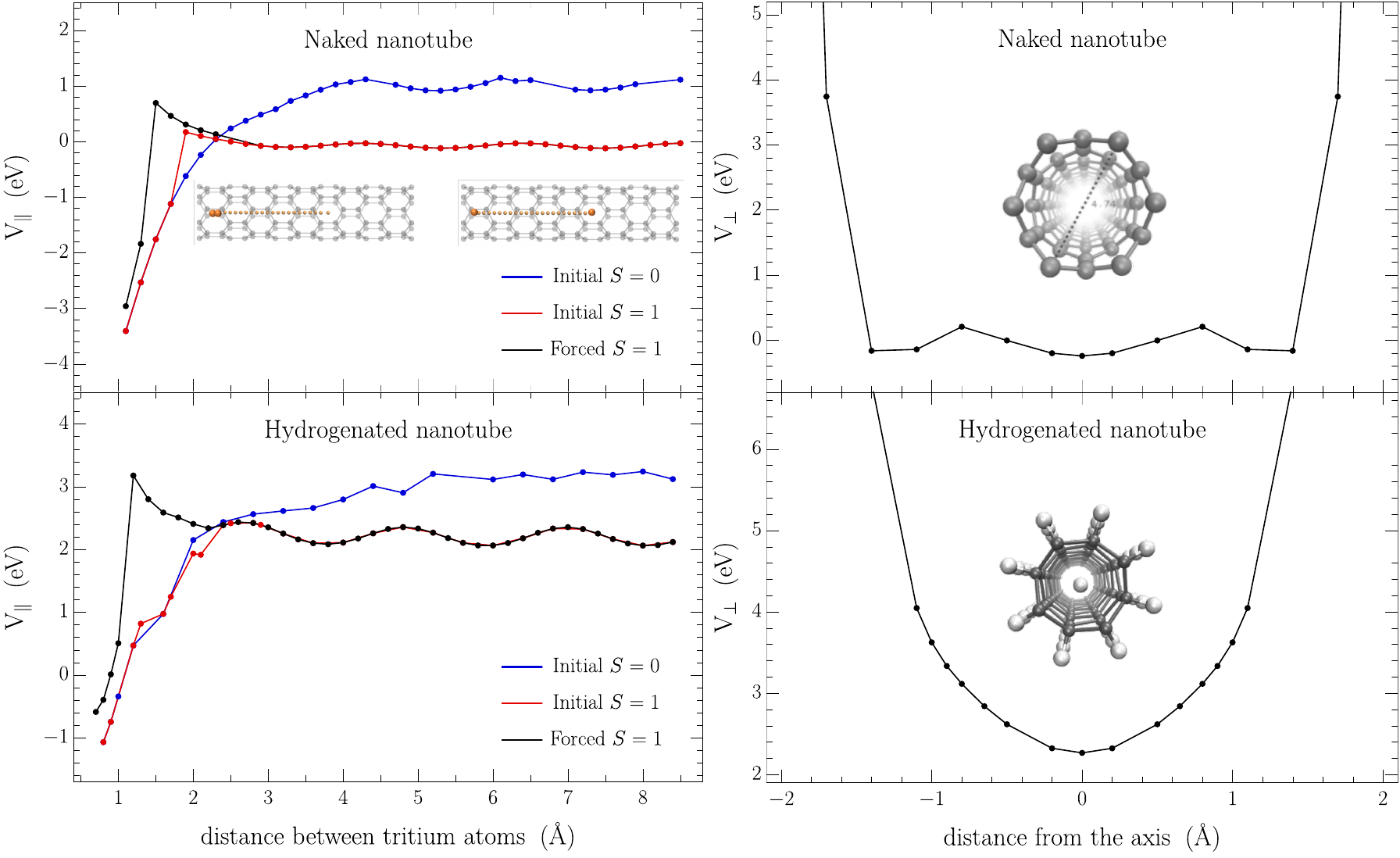} 
    \caption{Potential for a hydrogen atom inside a carbon nanotube. \textbf{Left panels:} Potential along the tube axis as a function of the relative distance between two tritium atoms (represented by the big orange dots in the insets) and for different spin configurations. When the atoms are well separated (rightmost inset) the potential is almost flat, with a small barrier depending on the spin configuration. When they get close to each other (leftmost inset) they tend to bind into a molecule, provided that their spin is (or can flip to) a singlet configuration (red and blue lines). If a triplet configuration is forced on the pair, a potential barrier prevents dimerization (black lines). In the lower panel, the hydrogenated nanotube potential shows a pronounced modulation because the radius of the hydrogenated nanotube is smaller than the bare one. In this case the hydrogen/tritium is affected by the atomic structure of the tube. This effect is less noticeable the greater the radius of the nanotube, as in the top panel. \textbf{Right panels:} Potential perpendicular to the tube axis. For a naked tube three minima are present, one at the center of the tube and two close to its walls. The latter correspond to the binding of the atoms to one of the carbon sites, which would prevent its motion along the axis. If the tube is hydrogenated, only the central minimum is left. Illustrative structures are reported in the insets. All energy profiles are measured with respect to gaseous atomic hydrogen.}
    \label{fig:nanotubes}
\end{figure*}


\subsection{Tritium in nanotubes} \label{sec:nanotubes}

The ideal setup to try and detect the cosmic neutrino background is one where the initial atomic tritium wave function is an eigenstate of momentum, as it would happen in vacuum. In this case no intrinsic quantum effects contribute to the uncertainty on the electron energy, which is then dominated by the experimental resolution. All the considerations reported in the previous sections indicate that flat graphene is not the optimal substrate to host the atomic tritium while still hoping to detect the neutrino background. However, the modulation of binding potentials and barriers operated by the local puckering and curvature of the sheet specifically suggests that concave sites might be favorable conformations to realize the desired almost-free tritium state. In this section we propose one possible configuration: the interior of carbon nanotubes.

To check the feasibility of this proposal, we used DFT to evaluate the potential felt by a hydrogen atom inside a nanotube of diameter between 4~\angstrom and 5~\angstrom, the smallest synthesized so far~\cite{smalltube}, as reported in  Figure~\ref{fig:nanotubes}. For the bare case, we studied a nanotube with a diameter of $4.8$~\angstrom and we find that the atom is almost completely free to move along the axis of the tube, with a potential that is essentially flat with weak modulations coming from roughly $70$~meV barriers. However, the inspection of the energy profile orthogonal to the axis reveals that, beside the central minimum where the hydrogen is almost free to slide in the center of the nanotube, there is a second minimum  corresponding to a configuration where the atom is bound to the internal surface of the tube. 

The possibility for the atom to bind to the tube is clearly not ideal, as it would prevent its free motion along the axis. This can be prevented by passivating the carbon nanotube, for example, with hydrogen bound to the external surface. In this case, due to the increase in the number of atoms, we study a nanotube with a diameter of $3.7$~\angstrom. Indeed, we find that, for a tube passivated in this way, the potential along the direction perpendicular to the tube does not feature the minima near the walls anymore---see, again, Figure~\ref{fig:nanotubes}. We also observe an increase in the periodic modulation of the potential along the tube axis, simply due to the smaller size of the tube. Larger tubes, as those realized in lab, should feature an essentially flat potential at large separations. Therefore, for single tritium atoms in a nanotube, this would realize the almost ideal situation of a free motion, at least in one dimension.

Nonetheless, if more than one tritium atom is present in each nanotube, the recombination is not completely prevented: the energy profiles clearly show a marked potential well for the formation of the molecule as the two atoms get close. However, the spin resolved calculations show that the electronic spin ground state of the molecule is the singlet, with null magnetization, while the electronic spin ground state of the separated atoms is the triplet. Indeed, when the initial spin configuration of the pair is $S=1$, there is a small barrier that must be overcome in order to form a molecule. In particular, if the pair is forced to be in a triplet configuration all the way, our calculation shows the emergence of a $\sim 1$~eV barrier preventing recombination. In the next section we discuss how this mechanism could be implemented to prevent dimerization of the \TT.

Before that, let us give a rough estimate of the amount of tritium that could be stored in carbon nanotubes.
Assuming that a recombination barrier has been achieved, we see from Figure~\ref{fig:nanotubes} that the minimum distance between tritium atoms is $\sim 3$~\angstrom, leading to a stoichiometric ratio of $\text{C}\!:\!\TT$ of at least $10\!:\!1$ or $20\!:\!1$ (depending on the tube). This implies a gravimetric loading that is an order of magnitude smaller than for graphene, i.e. between 10 and 20 g of tritium per kg of material.

\subsection{Magnetic fields to prevent dimerization} \label{ssec:B}

The peculiar spin ground state of the molecular hydrogen (and tritium) offers a way to possibly prevent the recombination of two atoms. As mentioned, the molecular hydrogen ground state is an electronic spin singlet ($S=0$), while the preferred spin state for atoms far apart is the triplet ($S=1$). It follows that a sufficiently high external magnetic field could force the pairs to be in spin triplet, hence preventing the molecular binding. Indeed, it has already been shown that \emph{in vacuum} an external field of $4-5$~T was capable of stabilizing atomic hydrogen at the temperature a few~Kelvin~\cite{newjphys2018,PRB88}. Moreover, the low dimensionality of our systems might help stabilizing the atomic tritium against dimerization. In fact, the recombination happening when the barrier is overcome is likely to be a nonadiabatic process happening through a spin state transition and the restrain to move in a single direction might reduce the accessible pathways to this process, consequently reducing its occurrence probability. As a consequence, this should increase the life time of the atomic tritium.
Recall that, indeed, when the hydrogen energy profile along the nanotube is evaluated forcing a spin triplet configuration, hence somewhat emulating the effect of a magnetic field, a barrier disfavoring dimerization appears (see again previous Section and Figure~\ref{fig:nanotubes}).

We further note that, by conservation of angular momentum, the introduction of a magnetic field parallel to the axis of the nanotube would induce a net polarization of the triutium sample, thus favoring the emission of electrons along the axis. This in turn would increase the favorable event rate.

It is also worth mentioning that while the benefit of having a magnetic field preventing, to some extent, the dimerization of tritium requires a quantitative description of the phenomenon, the PTOLEMY detector concept naturally foresees a magnetic field in the region of the target. 
The new electromagnetic filter design~\cite{Apponi:2021hdu}, on which the PTOLEMY detector concept is based, relies on the presence of a strong field, around 1~T, in the central region of the apparatus.
A continuously varying magnetic field is, in fact, crucial to allow for the measurement of the electron energy, as well as to guide it from the emission point to the final detector. The target region is, therefore, naturally immersed in such a field, which can be optimized according to the needs.

Finally, the recombination probability also depends on the concentration of hydrogen (tritium) in the system (surface or tube) and its mobility, both of which should be kept low to increase the half life of the atomic state. Clearly, these needs conflict with other needs in the PTOLEMY experiment: high tritium concentration to improve the event rate, and high tritium mobility to make it as close as possible to a momentum eigenstate. Therefore, a delicate balance between these parameters is needed, which could be achieved by properly choosing the material and the environmental conditions.


\section{Conclusion}

The localization of the initial tritium atom on the graphene sheet induces an intrinsic quantum spread in the energy spectrum of the electron emitted following either $\beta$-decay or neutrino capture. For the simplest case of a single tritium atom bound to a flat graphene sheet, this uncertainty is predicted to be at least an order of magnitude larger than the energy resolution expected in the PTOLEMY experiment~\cite{Cheipesh:2021fmg,Nussinov:2021zrj}.

In this work we determine quantitatively the expected rate as a function of the electron kinetic energy, under a set of simplifying assumptions. Building on what was highlighted in~\cite{Cheipesh:2021fmg}, we explain the important role played by the fate of the final \He atom, in particular whether it ends up in a continuous or discrete state of the potential. In the first case, the neutrino capture peak is hidden under the $\beta$-decay part of the spectrum. In the second case, instead, and when no additional degrees of freedom are excited, the neutrino capture peak remains well separated from the $\beta$-decay continuum, but its rate is exponentially suppressed, making this event highly unlikely.

Nevertheless, the possibility of final states where the \He atom remains bound to the graphene allows for electron energies up to a few electronvolts higher than in vacuum. Interestingly, this could be used as an experimental signature: in a setup like the one proposed by PTOLEMY one should observe electrons considerably more energetic than in vacuum.

We also perform a careful study of the tritium--graphene potential, and especially of its dependence on coverage as well as local geometry of the sheet. We propose carbon nanotubes as a possible solution to reduce the intrinsic quantum uncertainty. Inside a carbon nanotube passivated with hydrogen a tritium atom would be free to slide along the axis, hence realizing in one dimension a situation analogue to what happens in vacuum. Moreover, the introduction of an external magnetic field, which forces a pair of tritium atoms in a triplet configuration, might be able to prevent the formation of molecules.

Nonetheless, a number of effects could still affect the final electron shape. In particular, we expect initial state thermal motion to be a source of uncertainty on the electron energy, as well as the possible excitation of further degrees of freedom in the final state, like vibrational modes and nearby electrons. The role played by the latter has been discussed qualitatively in~\cite{Nussinov:2021zrj}, and more quantitatively in~\cite{Tan:2022eke}. It has been shown that the excitation of electronic degrees of freedom in the final state likely leads to further distortion in the final spectrum. Nonetheless, the analysis of~\cite{Tan:2022eke} shows that such an effect does not cause the disappearing of the absorption peak, which remains a qualitative feature of the spectrum. On top of this, since the $Q$-value for the atomic tritium is different from that of the molecular one, the occasional (although rare) dimerization would also induce additional uncertainty on the electron energy. A detailed, quantitative analysis of these (and other) effects and how to account for them is beyond the scope of the present study, and we leave it for future work.

As far as carbon nanotubes are concerned, we instead remark that they can be synthesized by growing them with a very good degree of parallelism using, for instance, a chemical vapor deposition technique. They can be arranged to be vertically aligned  and orthogonal to a  proper substrate (silicon, fused silica, stainless steel). Carbon nanotubes few hundreds of micrometer tall can be easily produced in relatively large quantities (10 mg/cm$^2$)~\cite{Apponi:2021lyd,Cavoto:2019flp,https://doi.org/10.1002/adbi.201800286}. There are currently ongoing studies to determine the level of hydrogenation of the nanotubes, and the possibility of storing tritium in them.

In conclusion, the outcome of this paper points to a possible spread in the energy of the electron emerging from the $\beta$-processes, due to an intrinsic quantum mechanical feature of the tritium--graphene bound state. The amplitude of the effect, however, needs to be directly measured. The PTOLEMY project aims at precisely addressing this issue by using the detector prototype~\cite{Apponi:2021hdu}, currently under construction at the Laboratory Nazionali del Gran Sasso. The prototype will be capable to reach an electron energy resolution of $50-100$~meV well below the size of the aforementioned effect.
To react to a possible confirmation of the phenomenon discussed in this paper the PTOLEMY collaboration has already started investigating other possible substrates, which will allow to overcome the limitation of graphene.


\begin{acknowledgments}
We are grateful to N.~Arkani-Hamed, J.~Maldacena, A.~Pilloni and R.~Rattazzi for important discussions. A.E.~is a Roger Dashen Member at the Institute for Advanced Study, whose work is also supported by the U.S.~Department of Energy, Office of Science, Office of High Energy Physics under Award No. DE-SC0009988. G.M. acknowledges the IT center of the University of Pisa, the HPC center (Franklin) of the IIT of Genova, and the allocation of computer resources from CINECA, through the ISCRA C projects HP10C9JF51, HP10CI1LTC, HP10CY46PW, HP10CMQ8ZK. G.M. also acknowledges support from the University of Pisa under the “PRA - Progetti di Ricerca di Ateneo” (Institutional Research Grants) - Project No. PRA 2020-2021 92. The research leading to these results has also received funding from the European Union’s Horizon 2020 research and innovation program under grant agreement no. 881603-Graphene Core3. T.F. acknowledges Funda\c{c}ao de Amparo \`a Pesquisa do Estado de Sao Paulo (FAPESP)  grants 2017/05660-0 and 2019/07767-1, and  Conselho Nacional de Desenvolvimento Cient\'\i fico e Tecnol\'ogico (CNPq) grants 308486/2015-3  and 464898/2014-5 (INCT-FNA).
Support and computational resources are also provided by EU under FETPROACT LESGO (Agreement No. 952068) and by the Italian University and Research Ministry, MIUR under MONSTRE-2D PRIN2017 KFMJ8E. 
We received support from the Amaldi Research Center funded by the MIUR program ``Dipartimento di Eccellenza'' (CUP:B81I18001170001). 
This work is further supported by the Italian grant 2017W4HA7S ``NAT-NET: Neutrino and Astroparticle Theory Network'' (PRIN 2017) funded by the Italian Ministero dell’Istruzione, dell’Universit\`a e della Ricerca (MIUR).
\end{acknowledgments}


\appendix

\section{Event rates} \label{app:rates}

Let us show concretely how the arguments of Section~\ref{sec:nutshell} come about. For the sake of the present work, it is enough to consider the case of a single neutrino. Let us start by comparing the rate obtained when the \He is freed from the graphene to that obtained when it remains in the ground state of the binding potential.
For simplicity we assume an isotropic wave function for the ground state,
\begin{align}
    \psi_i(\bm{x}) = \frac{1}{\pi^{3/4}\lambda^{3/2}} e^{-\frac{x^2}{2\lambda^2}}\,.
\end{align}
The width is taken from the harmonic approximation of the graphene potential at maximum coverage (i.e., graphane, see Figure~\ref{fig:potentials}) around its minimum, $\lambda \equiv (m \kappa)^{-1/4}$, with $m$ the atomic \TT or \He mass, and $\kappa \simeq 29$~eV/$\angstrom^2$ is the fitted spring constant. When the helium remains in the ground state, the matrix element is
\begin{align}
    \mathcal{M}_{\beta,\text{gs}} = \frac{g}{V} e^{-\lambda^2 |\bm{k}_\beta+\bm{k}_\nu|^2/4}\,,
\end{align}
and the corresponding event rate---i.e. the transition probability rate times the number of tritium atoms---is
\begin{widetext}
\begin{align}
\begin{split}
    dR_{\beta,\text{gs}} &= N_\text{T} \frac{|g|^2}{V^2} e^{-\lambda^2|\bm{k}_\beta+\bm{k}_\nu|^2/2} (2\pi) \delta(m_\TT - m_\He - E_\beta - E_\nu) \frac{V d^3k_\beta}{(2\pi)^3} \frac{V d^3k_\nu}{(2\pi)^3} \\
    &= N_\text{T} \frac{|g|^2}{(2\pi)^5} e^{-\lambda^2|\bm{k}_\beta+\bm{k}_\nu|^2/2} k_\nu E_\nu d^3k_\beta d\Omega_\nu\,,
\end{split}
\end{align}
\end{widetext}
with $N_\text{T}$ the number of tritium atoms, and $\Omega_\nu$ the solid angle of the outgoing neutrino.
In the second equality we have integrated over the neutrino energy using the $\delta$-function. Focusing on the near-endpoint region, we can now neglect the neutrino momentum in the exponential and perform the final integrals, obtaining
\begin{align}
    \frac{dR_{\beta,\text{gs}}}{dE_\beta} = N_\text{T}\frac{|g|^2}{2\pi^3}e^{-\lambda^2k_\beta^2/2} k_\nu E_\nu k_\beta E_\beta\,.
\end{align}
The maximum electron energy for this final state is $E_\beta^\text{max} = Q + m_e - m_\nu$. Note, importantly, that here we are considering the graphene sheet as infinitely massive, and therefore neglecting its recoil energy. This corresponds to the event where no vibrational modes are excited.

When the helium is, instead, freed from the graphene, its wave function is a plane wave. The matrix element is
\begin{align}
    \mathcal{M}_{\beta,\text{f}} = \frac{g}{V^{3/2}} 2^{3/2} \pi^{3/4} \lambda^{3/2}e^{-\lambda^2|\bm{k}_\text{He} + \bm{k}_\beta + \bm{k}_\nu|^2/2}\,,
\end{align}
and the event rate is given by
\begin{widetext}
\begin{align}
    dR_{\beta,\text{f}} &= N_\text{T}\frac{|g|^2}{V^3} 2^3 \lambda^3 \pi^{3/2} e^{-\lambda^2|\bm{k}_\text{He} + \bm{k}_\beta + \bm{k}_\nu|^2} (2\pi) \delta\left(m_\TT - \varepsilon_0 - m_\He - \frac{k_\text{He}^2}{2m_\He} - E_\beta - E_\nu\right) \frac{V d^3k_\text{He}}{(2\pi)^3} \frac{V d^3k_\beta}{(2\pi)^3} \frac{V d^3k_\nu}{(2\pi)^3} \notag \\
    & = N_\text{T}\frac{|g|^2}{32\pi^{13/2}} \lambda^3 e^{-\lambda^2|\bm{k}_\text{He} + \bm{k}_\beta + \bm{k}_\nu|^2} d^3k_\text{He} k_\nu E_\nu d^3k_\beta d\Omega_\nu\,,
\end{align}
\end{widetext}
where again we used conservation of energy to fix the energy of the neutrino. Here $\varepsilon_0 = U_0 - \frac{3}{2}\sqrt{\kappa/m}\simeq 5.76$~eV is the ground state binding energy. Now we again neglect the neutrino momentum in the exponent and, integrating over the electron and helium solid angles, we obtain
\begin{align}
    \begin{split}
        \frac{dR_{\beta,\text{f}}}{dE_\beta} &= N_\text{T} \frac{|g|^2 \lambda}{2\pi^{7/2}}E_\beta \\
        &\quad \times \int_0^{k_\text{He}^\text{max}} dk_\text{He} \, e^{-\lambda^2(k_\text{He}-k_\beta)^2} k_\text{He} k_\nu E_\nu\,, 
    \end{split}
\end{align}
where we have neglected exponentially small terms in the integral. The maximum helium momentum is obtained requiring that $E_\nu \geq m_\nu$, which returns $k_\text{He}^\text{max} = \sqrt{2m_\He(Q+m_e-m_\nu - \varepsilon_0 - E_\beta)}$. Correspondingly, the maximum energy that the electron can have in this configuration is $Q+m_e-m_\nu-\varepsilon_0$.

The rates for neutrino absorption are given by
\begin{align}
    \frac{dR_\text{CNB}}{dE_\beta} = \frac{d\sigma_\text{CNB}}{dE_\beta} N_\text{T} n_\nu v_\nu f_c\,,
\end{align}
where $n_\nu \simeq 56$~cm$^{-3}$ is the neutrino density, while $f_c$ is the so-called clustering factor. The latter satisfies $f_c\geq 1$, and for sufficiently large masses follows the empirical law, $f_c\simeq 76.5\,(m_\nu/\text{eV})^{2.21}$~\cite{PTOLEMY:2019hkd}. The absorption cross section, $\sigma_\text{CNB}$, can be found in similar ways as shown above, and we will not report the details here. 

The reason why the absorption peak for a free helium ends up below the $\beta$-decay spectrum is that, while the maximum energy allowed for the absorption is still larger than the maximum energy for the $\beta$-decay by $2m_\nu$, the most likely one happens when the $k_\text{He} = k_\beta$, which gives an energy that, compared to the maximum energy for the $\beta$-decay in this final state, is smaller by roughly $m_e Q/m_\He\simeq 3.4$~eV.
We also notice that, the only instance where the absorption peak is well separated from the rest of the spectrum is the case when the electron has the maximum allowed energy, i.e. when the helium remains bound in its ground state. This event, however, happens with very small probability.

\section{Recoil of heavy and light degrees of freedom} \label{app:recoil}
Here we show how other degrees of freedom can recoil after the emission of the electron. 
We do that with a simple toy example: a free tritium atom. We denote as $\bm{R}$ the center of mass coordinate and as $\bm{r}$ the relative distance of the atomic electron from the decaying/decayed nucleus. The initial wave function will be given by 
\begin{align}
    \psi_i(\bm{R},\bm{r}) \propto \phi_0^{(Z=1)}(\bm{r})\,,
\end{align}
where $\phi_0^{(Z=1)}$ is the ground state hydrogenic wave function. We also took the (free) center of mass to be at rest. After the $\beta$-decay, the final wave function will be
\begin{align}
    \psi_f(\bm{R},\bm{r},\bm{x}_\beta) \propto e^{i \bm{P}\cdot\bm{R}} e^{i \bm{k}_\beta\cdot\bm{x}_\beta} \phi_n^{(Z=2)}(\bm{r})\,,
\end{align}
where we neglect the neutrino momentum. Here $\phi_n^{(Z=2)}$ is the wave function for some excited state of the hydrogenic atom with two positive charges.
The weak matrix element forces the $\beta$ electron to be produced in the same location as the initial tritium and final helium. The matrix element then reads
\begin{widetext}
\begin{align}
\begin{split}
    \mathcal{M}_{fi} &\propto \int d^3R \, d^3r \,e^{-i(\bm{P}+\bm{k}_\beta)\cdot\bm{R}}\, \phi_0^{(Z=1)}(\bm{r})\,\phi_n^{(Z=2)}(\bm{r}) \, e^{-i \frac{m_e}{m_e+m_\He}  \bm{k}_\beta\cdot\bm{r}} \\
    &\propto \delta(\bm{P}+\bm{k}_\beta) \int d^3r \,\phi_0^{(Z=1)}(\bm{r})\,\phi_n^{(Z=2)}(\bm{r})\,e^{-i \frac{m_e}{m_e+m_\He}  \bm{k}_\beta\cdot\bm{r}}\,.
\end{split}
\end{align}
\end{widetext}
where we used the fact that, since the $\beta$ electron is on the same position as the helium nucleus, we can write $\bm{x}_\beta = \bm{R}+\frac{m_e}{m_e+m_\He}\bm{r}$. 

The equation above is telling us that, on the one hand, momentum is conserved by the recoil of the system as a whole. On the other hand, since the hydrogenic wave functions are localized over a distance of the order of the Bohr radius, $a$, while $\frac{m_e}{m_e+m_\He}k_\beta a\ll 1$, the atomic electron can be excited to some higher discrete level while still be bound---the matrix element does not suffer from the exponential suppression discussed in Section~\ref{sec:nutshell}.


\section{Analytic form of the binding profiles} \label{app:valapp}

\begin{figure}[t]
    \centering
    \includegraphics[width=0.3\textwidth]{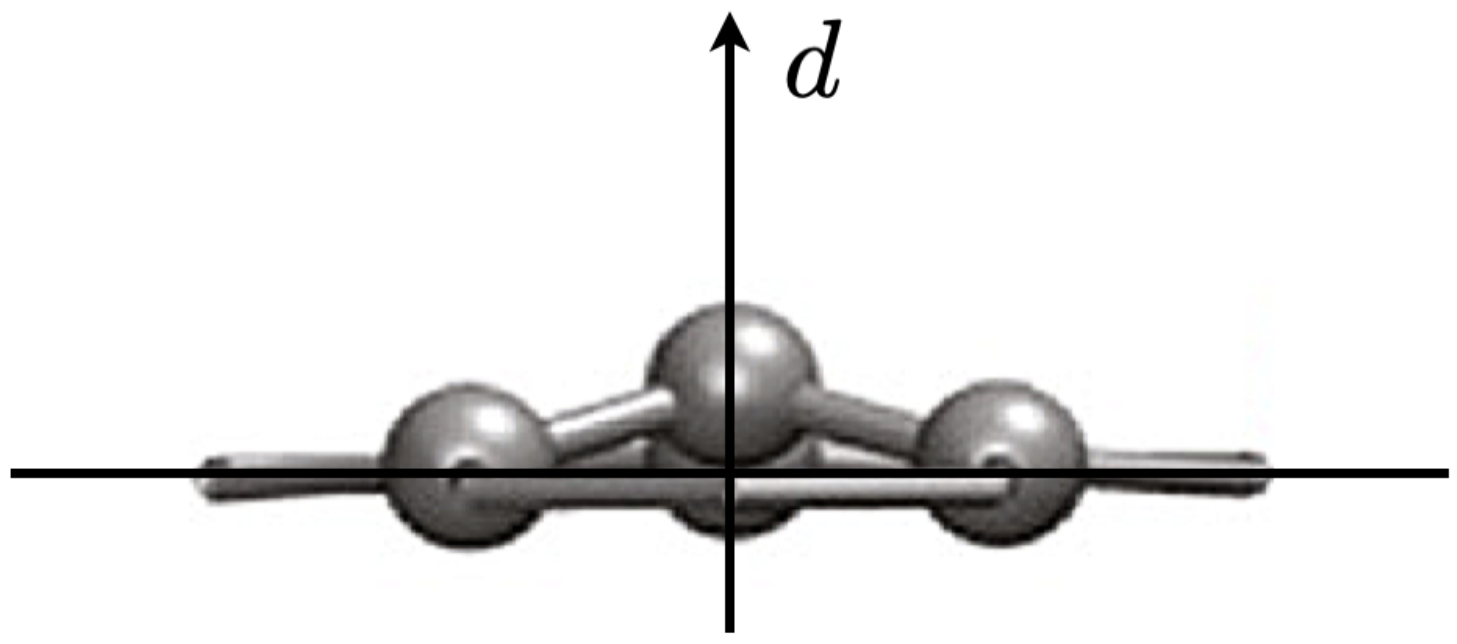}
    \caption{Definition of the puckering distance $d$, i.e.\ the out-of-plane displacement of a carbon atom with respect to plane defined by its three neighbors, here used as a measure of the local curvature of the sheet. We indicate with $d>0$ convexities and with $d<0$ concavities.}
    \label{fig:pucker}
\end{figure}

The analytical H--C binding profile of Figure~\ref{fig:potentials}(a) is obtained combining two Morse potentials, one for the chemical binding and one for the van der Waals interaction, in a way such that the former dominates at small distances and the latter at large distances:
\begin{align}
    V(r) = u_\text{ch}(r)f(r) + u_\text{vdW}\big(1-f(r)\big)\,,
\end{align}
with
\begin{align}
    u_\text{ch}(r) &= \big(\varepsilon_\text{b}+\varepsilon_\text{off}\big)\left[ \left( e^{-\frac{r-r_\text{ch}}{\alpha_\text{ch}}} -1 \right)^2 -1 \right] + \varepsilon_\text{off} \,, \notag \\
    u_\text{vdW}(r) &= \varepsilon_\text{vdW}\left[ \left( e^{-\frac{r-r_\text{vdW}}{\alpha_\text{vdW}}} -1 \right)^2 -1 \right]\,, \\
    f(r) &= \frac{1}{e^\frac{r-R}{\sigma}+1}\,. \notag
\end{align}
Here $r_\text{ch}\simeq1.1$~\angstrom and $r_\text{vdW}\simeq2.7$~\angstrom are respectively the chemical and van der Waals C--H distances, while the chemical and van der Waals widths are $\alpha_\text{ch}\simeq0.2$~\angstrom and $\alpha_\text{vdW}\simeq1.5$~\angstrom. Moreover, while the depth of the van der Waals potential is fixed to be $\varepsilon_\text{vdW}\simeq 6$~meV, the one of the chemical potential is a function of the puckering distances, $d$ (see Figure~\ref{fig:pucker}), specifically $\varepsilon_\text{b} \simeq \big( 0.8 + 4.45 (d/\angstrom) \big)$~eV and $\varepsilon_\text{off}\simeq \big( 0.5 - 0.2 (d/\angstrom) - 2.5 (d/\angstrom)^2\big)$~eV. These expressions are obtained by fitting the dependence of the barrier height and of the hydrogen binding energy on the curvature from~\cite{valjcpc2011}. The parameters of the function $f(r)$ are tuned to reproduce the correct barrier height for flat graphene, and are given by $\sigma\simeq0.15$~\angstrom and $R\simeq2.0$~\angstrom.

Note that from this one can compute the value of the spring constant corresponding to the minimum of the potential, obtaining
\begin{align}
    \kappa \simeq 2(\varepsilon_\text{b}+\varepsilon_\text{off})/\alpha^2_\text{ch}\,.
\end{align}
This can be strongly reduced for large values of negative $d$, i.e. within concavities.

\section{Density Functional Theory calculation} \label{app:dft}
We carried out the DFT calculations 
with \textsc{Quantum Espresso}~\cite{QE1,QE2,QE3}, which uses a plane wave basis set. The pseudopotentials were taken from the standard solid-state pseudopotential efficiency library~\cite{Prandini_2018,Lejaeghere_2016,Schlipf_2015,Dalcorso2014,Garrity_2014} with cutoffs of 60 Ry and 480 Ry for the wave functions and the density. The exchange-correlation potential was treated
in the Generalized Gradient Approximation, as parametrized by the Perdew--Burke--Ernzerhof formula~\cite{PBE}, with the van der Waals (vdW)-D2 correction as proposed by Grimme~\cite{Grimme_2006}. For the Brillouin Zone integrations, we employed a Marzari--Vanderbilt smearing~\cite{Marzari_1999} of $4\times 10^{-3}~{\rm Ry}$ with a Monkhorst--Pack~\cite{Monkhorst_1976} k-point grid with $12\times 12\times 1$ points for self-consistent calculations of the charge density and for geometry optimization of the graphene sheets. For the carbon nanotubes ((0,6) naked and (0,4) H-passivated) we used supercells with the exact periodicity of 5 repeated units along the $z$-axis and $\Gamma$ point self-consistent calculations.  We optimized the geometrical structures relaxing the atomic positions until the components of all the forces on the ions are less than $10^{-3}~{\rm Ry/Bohr}$ (the supercells of the nanotubes were also relaxed along the direction of the axis of the tube). For the 2D graphene-like geometries we employed  $4\times 4$ and $5\times 5$ supercells, starting with a unit cell with a lattice constant $a_0 = 2.46$~\angstrom~\cite{Grosso,Colle_2016}. We considered a supercell with about 18~\angstrom of vacuum along the orthogonal direction to avoid interaction between the periodic images. Analogously, we left approximately 12~\angstrom of lateral free space between nanotubes. 
Besides spin-unrestrained calculations, in order to mimic the effect of an external magnetic field, we evaluated the energy profile of the hydrogen atom along the carbon nanotube adding a penalty function to the total energy to restrain the spin configurations to the initial ones (singlet or triplet).

The minimum energy paths of Figure~\ref{fig:hop} have been obtained  using the climbing image version~\cite{Henkelman_2000,Colle_2016} of the Nudget Elastic Band (NEB) method as implemented in the \textsc{QE} code. This method is a modification of the regular NEB method~\cite{Johannesson_1998,Henkelman_2002} to converge rigorously to the highest saddle point (transition state) on the energy surface containing the initial and the final chosen states. 

We use the VESTA~\cite{vesta}, Xcrysden~\cite{xcrys} and VMD\cite{vmd} graphics software tools to visualize the geometrical structure, to produce the plots and to generate the starting structures of the nanotubes.

\bibliography{bibliography}

\end{document}